\documentclass[aps,amsmath,amssymb,prb,twocolumn,showpacs]{revtex4}
\usepackage{bm}
\usepackage{epsfig}

\newcommand{\coma}{cond-mat/}

\newcommand{\etal}{{\textit{et al.}}}

\newcommand{\di}{\textrm{d}}

\newcommand{\dis}{\Delta}

\newcommand{\ba}{{\mathbf a}} 

\newcommand{\br}{{\mathbf r}}

\newcommand{\grad}{{\bm \nabla}}

\newcommand{\cH}{{\mathcal H}}
\newcommand{\cS}{{\mathcal S}}
\newcommand{\cD}{{\mathcal D}}
\newcommand{\cHf}{{\mathcal H}_\textrm{free}}
\newcommand{\cHi}{{\mathcal H}_\textrm{int}}

\newcommand{\ve}{{\vec e}}
\newcommand{\vf}{{\vec f}}
\newcommand{\vS}{{\vec S}}
\newcommand{\vs}{{\vec s}}
\newcommand{\vmu}{{\vec \mu}}

\newcommand{\hR}{{\hat R}}

\newcommand{\tp}{{\widetilde p}}
\newcommand{\tJ}{{\widetilde J}}
\newcommand{\tdis}{{\widetilde \dis}}

\newcommand{\ellcr}{{\ell^\perp}}

\newcommand{\0}{{(0)}}
\newcommand{\1}{{(1)}}

\begin{document}

\title{Bond-disordered spin systems:  
Theory and application to doped high-$T_\textrm{c}$ compounds}

\author{Frank Kr\"uger}
\author{Stefan Scheidl}

\affiliation{Institut f\"ur Theoretische Physik, Universit\"at zu
  K\"oln, Z\"ulpicher Str. 77, D-50937 K\"oln, Germany}

\date{\today}

\begin{abstract}
  We examine the stability of magnetic order in a classical Heisenberg
  model with quenched random exchange couplings. This system
  represents the spin degrees of freedom in high-$T_\textrm{c}$
  compounds with immobile dopants.  Starting from a replica
  representation of the nonlinear $\sigma$-model, we perform a
  renormalization-group analysis.  The importance of cumulants of the
  disorder distribution to arbitrarily high orders necessitates a
  functional renormalization scheme.  From the renormalization flow
  equations we determine the magnetic correlation length numerically
  as a function of the impurity concentration and of temperature.
  From our analysis follows that two-dimensional layers can be
  magnetically ordered for arbitrarily strong but sufficiently diluted
  defects.  We further consider the dimensional crossover in a stack
  of weakly coupled layers.  The resulting phase diagram is compared
  with experimental data for La$_{2-x}$Sr$_x$CuO$_4$.
\end{abstract}

\maketitle


\section{introduction}

Although the interest in disordered spin systems reaches back several
decades (for a review, see e.g. Ref. \onlinecite{Binder+86}), interest
has strongly revived in recent years by the recognition that
high-$T_\textrm{c}$ compounds exhibit phases with antiferromagnetic
(AFM) order, spin-glass order or stripe order in certain ranges of
temperature and doping.  A prominent example for such materials is
La$_{2-x}$Sr$_x$CuO$_4$, in which every Cu atom carries a spin $\frac
12$ (for an overview on this material, see e.g. Ref.
\onlinecite{Shirane+94}).

In these materials, the spins are located in weakly coupled layers.
Within each layer, the spins constitute a square lattice with
antiferromagnetic exchange coupling between nearest
neighbors.\cite{note:struct} In the undoped system, these spins can be
represented to a good approximation by a classical two-dimensional
model since quantum fluctuations lead to a merely quantitative
renormalization of the classical parameters.\cite{Chakravarty+89}
Although a weak coupling between the layers and a weak easy-plane
anisotropy are present in these materials (both are about five orders
of magnitude smaller than the isotropic intra-plane
exchange\cite{Keimer+92a}), they become relevant only on relatively
large scales.  Thus, on the finite length scales of experimental
relevance, the spins can be described in a first approximation by a
classical Heisenberg antiferromagnet in two dimensions.

Doping induces holes in the layers, which can lead to quenched
frustration of the exchange couplings.  Frustration can occur when the
holes are localized individually\cite{Aharony+88,Gooding+91} as well
as when they condense into a topologically defective array of stripes
which act as antiphase domain walls for the AFM
order.\cite{Zaanen+01,Hasselmann+00} Both cases can be represented by
an effective bond-disordered Heisenberg model. Such a system is
expected to display generic spin-glass behavior, as is observed in
experiments beyond a critical doping.\cite{Chou+95}

On the theoretical side, Heisenberg spin glasses are much less
understood than Ising spin glasses.  While it is well established for
the pure two-dimensional system that the magnetic correlation length
$\xi$ decays exponentially with increasing
temperature,\cite{Polyakov75} the dependence of $\xi$ on disorder is
controversial.  At $T=0$ it is not clear whether $\xi$ is finite for
every small but finite concentration of defect
bonds\cite{Villain79,Glazman+90,Shirane+94,Cherepanov+99} or whether
$\xi$ is infinite up to a critical concentration even for strong
defects.\cite{Rodriguez+95} Concerning the temperature dependence of
$\xi$, there is no consensus as to whether it is
reentrant\cite{Glazman+90} or not.\cite{Cherepanov+99} To readdress
these unsettled issues, we develop a theoretical approach for bond
disordered spin glasses which complements previous approaches.  In
particular, we calculate the correlation length and compare our
results to the results of previous theoretical approaches and to
experiments on high-$T_\textrm{c}$ compounds.

The outline of the paper is as follows. In Sec. \ref{sec.model} we
establish the model and briefly review the mechanism by which holes
generate magnetic textures that provide the nucleus for a reduction of
magnetic order.  Based on this mechanism, we further motivate why it
is desirable to develop an alternative to previous theoretical
approaches.  In Sec. \ref{sec.th}, we derive the functional
renormalization group equations.  These flow equations are evaluated
numerically and our results for the magnetic correlation length are
compared to experimental data in Sec. \ref{sec.res}, where we also
consider the dimensional crossover from 2D to 3D behavior.  We
conclude with a discussion of our results in Sec. \ref{sec.concl}.

\section{Model}
\label{sec.model}

In a system with sparse substitutional doping, the dopants can be
considered as randomly distributed with negligible correlations and as
quenched over a wide temperature range.  In doped Mott insulators, the
effect of the dopants is twofold: they induce holes in the cuprate
planes and at the same time they provide a random potential that
localizes these holes.  Therefore, they are commonly assumed to be
localized on the oxygen atoms between neighboring copper
atoms.\cite{Aharony+88,Emery+88,Gooding+91}

A simple argument suggests that such a hole transmutes the
antiferromagnetic superexchange coupling between the neighboring
copper spins into a ferromagnetic one.  Thereby frustration is induced
in the spin system.\cite{Aharony+88} The irrelevance of quantum
fluctuations of the spins in the pure system\cite{Chakravarty+89} (in
the sense that they lead only to a quantitative renormalization of
classical parameters such as the spin stiffness) suggests that they
may be neglected in a good approximation also for small
doping.\cite{note:QF}

\subsection{Definition}

Starting from a classical description of the AFM, we may examine an
equivalent model where all spins of one bipartite sublattice are
flipped and the sign of all exchange couplings is
reversed,\cite{Fradkin+78} i.e., we now consider the coupling of the
undoped system to be ferromagnetic and the defect couplings to be
antiferromagnetic.  We base our analysis on the Hamiltonian
\begin{eqnarray}
  H =  \frac 12  \sum_{\br,i}  J_i(\br)  \ [\nabla_i \vS (\br)]^2 
\label{H}
\end{eqnarray}
for spins on a (hyper)cubic lattice in $d$ dimensions (a single layer
is described by $d=2$).  Spins are treated as classical $N$-component
vectors of unit length, $\vS^2(\br)=1$.  A spin at site $\br$ is
coupled to its nearest neighbor in direction $i=1,\cdots, d$ via the
exchange coupling $J_i(\br)$.  We define $\nabla_i \vS (\br):= \vS
(\br+\ba_i) - \vS (\br)$ for a basis vector $\ba_i$.  The global
symmetry $O(N)$ of spin rotations is preserved for arbitrarily
disordered exchange couplings.

It is convenient to rewrite the exchange coupling as
\begin{eqnarray}
  J_i(\br) =   [1- \dis_i(\br)] J 
\end{eqnarray}
with the value $J>0$ of the pure system and the quenched random
variable $\dis_i(\br)$.  Frustration effects emerge for
$\dis_i(\br)>1$, when a ferromagnetic bond becomes antiferromagnetic.
To further specify the nature of disorder, we assume a bimodal
distribution of the exchange couplings,
\begin{eqnarray}
  \dis_i(\br)=
  \left\{ 
    \begin{array}{ll}
      \dis & \textrm{ with probability } p,
      \\
      0 & \textrm{ with probability } 1-p.
    \end{array}
  \right.
  \label{bimodal}
\end{eqnarray}
For square lattices as in La$_{2-x}$Sr$_x$CuO$_4$, the concentration
$x$ is related to this probability by $x=2p$.  Correlations between
different bonds are assumed to be absent.

\subsection{Single defects}

Before we address the magnetic order in the presence of a finite
concentration of defects, it is instructive to recall briefly the
physics of a \textit{single} defect for $N \geq 2$.  While the ground
state of the pure system is collinear (all spins are strictly
parallel), it is canted (i.e., no longer collinear) for a single
defect beyond a certain critical strength
$\dis_\textrm{single}$.\cite{Villain77a,Villain77b,Villain79} The
threshold $\dis_\textrm{single}=d$ value can be obtained from a
spin-wave calculation\cite{Saslow+88} (see also appendix
\ref{app.cont}). The ground state for a single defect with $\dis >
\dis_\textrm{single}$ remains \textit{planar} for all $N \geq 2$,
i.e., apart from a global rotation, the defect texture for Heisenberg
spins ($N=3$) is identical to the texture for XY spins ($N=2$).  Far
away from the defect bond, the texture is described by the solution of
the Laplace equation\cite{Villain79}
\begin{eqnarray}
  \grad^2 \vS (\br) = 0 
  \label{laplace}
\end{eqnarray}
and the spins approach a collinear configuration, $\vS(\br) \to \vS_0$
for $r \to \infty$.  To be specific, we discuss a defect bond located
at $\br=0$ and oriented in direction $\ba$.  It acts as a source of a
dipolar distortion with a moment $\vmu$ perpendicular to $\vS_0$.  For
$d=2$, in particular,
\begin{eqnarray}
  \vS(\br) - \vS_0 \approx \frac{\vmu}{2\pi}  
  \frac{\ba \cdot \br }{r^2} .
\end{eqnarray}
The amplitude of the dipole moment is determined by the nonlinearity
of the model, which may be considered as interaction between spin
waves.  The magnetic textures of several defect bonds are subject to a
dipolar interaction at large distances.\cite{Villain77a,Villain79} The
effects of a finite density of defects are nontrivial because of the
frustrated long-range interaction between the dipoles.  In the
presence of more than two defects, the interaction among the dipoles
typically becomes frustrated. Then the ground state is no longer
planar for $N>2$.

\subsection{Finite defect concentration}
\label{sec.finite}

Depending on the relative position of several defect bonds in a
cluster, the generated texture may have dipolar or higher-order
multipolar structure.  In contrast to electrodynamics, the multipolar
moments of defect textures are not additive since the spins act as a
nonlinear medium for the dipoles, giving rise to many-body
interactions.  In particular, clusters of defects can lead to canting
already for defect-bond strengths \textit{below}
$\dis_\textrm{single}$, i.e., in clusters of defects the threshold
strength of the defect bonds is reduced.  The threshold values of some
specific bond configurations in the XY model are given in Refs.
\onlinecite{Saslow+88} and \onlinecite{Gawiec+91} (cf. in particular
Table I in the latter one).  In general, threshold values in the XY
system are upper bounds for the thresholds in a Heisenberg system
since with increasing $N$ the spins have a larger space of canted
states which they may explore to minimize energy.

Thus, as soon as the defects are antiferromagnetic ($\dis>1$) they can
induce canted textures due to the presence of arbitrarily large
clusters.\cite{Villain79,Parker+88,Saslow+88,Gawiec+91} Hence the
ground state will be canted for \textit{every} $p>0$ and $\dis>1$
(assuming that the defect distribution has no pathological spatial
correlations).  Therefore one may expect a system with a certain
density of weak defects $1< \dis <\dis_\textrm{single}$ to be
qualitatively equivalent to a system with a (possibly much) smaller
concentration of strong defects $\dis >\dis_\textrm{single}$.

The presence of canting certainly implies a reduction of
magnetization. However, one cannot draw direct conclusions about the
range of magnetic order.  For a sufficiently high density of strong
defects, one certainly has short-range order.  For defects of a weaker
strength or a lower concentration, one can have also quasi-long-range
order (as in a XY model with quenched and uncorrelated
dipoles\cite{Gawiec+91,Nattermann+95}).  A phase with true long-range
order is also possible but unlikely since it would require a highly
ordered dipole configuration.

It is instructive to recall a ``duality'' relation\cite{Gawiec+93} in
the $(p,\dis)$ parameter space, since in a mixture of ferromagnetic
and antiferromagnetic bonds there are two ways to declare one type as
``regular'' bond and the other type as ``defect'' bond (see appendix
\ref{app.dual}).  This duality implies a relation
\begin{eqnarray}
  \dis_\textrm{AFM}(p)= \frac{\dis_\textrm{FM}(1-p)}
  {\dis_\textrm{FM}(1-p)-1}
\end{eqnarray}
between boundary lines $\dis_\textrm{FM}(p)$ and
$\dis_\textrm{AFM}(p)$ limiting regions with ferromagnetic order (for
$\dis < \dis_\textrm{FM}$) and antiferromagnetic order (for $\dis >
\dis_\textrm{AFM}$) in the $(p,\dis)$ plane. Since the duality
relation maps the region $p \dis >1$ onto $p \dis <1$ it is sufficient
to examine the latter region.

To perform first qualitative estimates, one may consider the
disorder-averaged exchange coupling, which is
\begin{eqnarray}
  \overline {J_i(\br)} = (1-p \dis) J
  \label{J.av}
\end{eqnarray}
for the bimodal distribution (\ref{bimodal}). Thus, there is a
tendency towards the formation of ferromagnetic order for $p\dis <1$
(where $\overline J >0$) and a tendency towards antiferromagnetic
order for $p \dis >1$ (where $\overline J <0$).  Certainly, the
presence of order requires more than such a tendency.  A further
minimum requirement should be that the relative fluctuations
\begin{eqnarray}
  \sigma^2 := \frac{\overline{J_i^2(\br)} - \overline{J_i(\br)}^2}
  {\overline{J_i(\br)}^2}  = \frac{p(1-p) \dis^2}{(1-p \dis)^2}
\end{eqnarray}
of the exchange couplings must be small.  Since $\dis>1$ and $p
\dis<1$ in the range of interest, $\sigma^2 \ll 1$ requires $p \dis^2
\ll 1$.  A very crude estimate of the boundary from $\sigma^ \simeq 1$
suggests
\begin{eqnarray}
  \dis_\textrm{FM}(p) \simeq p^{-1/2}.
  \label{est.FM}
\end{eqnarray}

For the special case $\dis=1$, the defect bonds have a vanishing
exchange coupling and the system is bond-diluted.  The presence of
magnetic order (in the sense of a finite magnetization) then requires
that a finite fraction of spins is connected by regular bonds.  This
is the case below the percolation transition, i.e. for $p < \frac 12$
in $d=2$.\cite{Sykes+64} Therefore one expects $\dis_\textrm{FM}(p)=1$
for $p \geq \frac12$.

\subsection{Previous work}
\label{sec.prev}

There are only a few approaches in the literature aiming at a more
sophisticated analysis, which we briefly summarize in order to
highlight our motivation to reconsider this problem in an alternative
way.  First, the coherent-potential approximation (CPA) provides a
simple self-consistent approach to determine an effective spin
stiffness for the random system.  For $N=2$, the CPA
yields\cite{Vannimenus+89} a transition line
$\dis_\textrm{FM}^\textrm{CPA}(p)$ smoothly interpolating between
$\dis_\textrm{FM}^\textrm{CPA}(0)=d$ (reflecting the canting threshold
for individual defect bonds) and $\dis_\textrm{FM}^\textrm{CPA}(p)=1$
for $p \geq \frac 12$ in $d=2$ and for $p \geq \frac 23$ in $d=3$.  In
$d=2$, the location of the percolation transition is captured exactly.
This has to be considered as a fortunate coincidence which is absent
in $d=3$, where the location is found only
approximately.\cite{Vannimenus+89} In the limit $N \to \infty$ the CPA
yields that the stability of order requires
$\sigma^2<1$,\cite{Rodriguez+95} which results in
$\dis_\textrm{FM}^\textrm{CPA}(p) \approx p^{-1/2}$ for small $p$ in
agreement with the naive estimate (\ref{est.FM}).

Unfortunately, the interpretation of the CPA in somewhat
ambiguous.\cite{note:CPA} According to its construction, the CPA
replaces the disordered system by a ferromagnetic one with a
homogeneous effective exchange constant. The strength of this
effective coupling is determined from a selfconsistency condition that
considers the fluctuations of a single bond. Thus, the CPA misses
effects of clusters of defect bonds.  The putative transition line is
defined as the border line up to which selfconsistent solutions exist.
A priori, this border line can be interpreted in two ways: either as
the onset of the canting instability or as the onset of short-range
magnetic order.  Therefore, the CPA can be only of qualitative use.

To gain insight into the nature of the ground state, Gawiec and
Grempel\cite{Gawiec+91,Gawiec+93} (GG) performed numerical studies of
the XY model in $d=2$.  Their data suggest a transition line
$\dis_\textrm{FM}^\textrm{GG}(p)$ between a phase with
quasi-long-range order and short-range order.  It starts from at the
canting instability $\dis_\textrm{FM}^\textrm{GG}(0) =
\dis_\textrm{single}=2$ and follows
$\dis_\textrm{FM}^\textrm{GG}(p)=1$ beyond the percolation transition.
Thus, the data suggest short-range order for an \textit{arbitrarily
  small} concentration of defects of a strength exceeding the canting
instability of single defects in agreement with CPA.  However, as
stated in Ref.  \onlinecite{Gawiec+91}, the question of whether one
can have order for $\dis>\dis_\textrm{single}$ at sufficiently small
$p$ remains far from settled due to finite-size effects.

For the Heisenberg model with the bimodal bond distribution, there are
only very few numerical studies.  For the special case of two
dimensions and $\dis=\dis_\textrm{single}=2$, Nonomura and
Ozeki\cite{Nonomura+95} postulate order for $p \lesssim 0.11$ from an
exact-diagonalization method for $S=1/2$ quantum spins.  This implies
an even larger stability of the classical model, and in particular
$\dis_\textrm{FM}(0)>2$. However, also their conclusion has to
considered with care because of finite-size effects.

Previous analytic approaches\cite{Glazman+90,Cherepanov+99} are
implicitly restricted to defect bond strengths exceeding the canting
threshold, $\dis \gg \dis_\textrm{single}$.  The spin system with
random bonds is \textit{replaced} by a spin system with homogeneous
bonds coupled to an additional canting field that also generates
dipolar spin textures.  In order to shed some light on the quality of
this replacement, we perform a Hubbard--Stratonovich transformation
introducing an auxiliary bivectorial field $\vf_i$ via
\begin{subequations}
  \label{hs-trafo}
  \begin{eqnarray}
    \exp(- H[\vS]/T) &=& \int \cD \{\vf\} \
    \exp[-\tilde H[\vS,\vf]/T] ,
  \end{eqnarray}
  with
  \begin{eqnarray}
    \tilde H[\vS,\vf] &=& J \sum_{\br,i} \bigg\{ \frac 12 [\nabla_i \vS
    (\br)]^2 - \vf_i(\br) \cdot \nabla_i \vS  (\br)
    \nonumber \\ &&
    + \frac 1{2\dis_i(\br)} \vf_i^2(\br) \bigg\} .
    \label{H.sf}
  \end{eqnarray}
\end{subequations}
In this representation, $\vS$ and $\vf_i$ are thermally fluctuating
variables, $\vS$ with spherical constraint, $\vf_i$ unconstrained.  In
the Hamiltonian (\ref{H.sf}) of the transformed system, the spins
interact directly via the homogeneous exchange coupling $J$.  In
addition, they couple also to the canting field $\vf_i$.  It is
precisely such a coupling that was considered in Refs.
\onlinecite{Glazman+90,Cherepanov+99}.

The Hubbard--Stratonovich transformation -- which yields an exact
representation of the original model (\ref{H}) -- shows that the field
$\vf_i$ has a \textit{local} selfinteraction potential
\begin{eqnarray}
  V_{ij} (\br,\br')=\delta_{ij} \delta_{\br,\br'} J/\dis_i(\br) .
\end{eqnarray}
On regular bonds with $\dis_i(\br)=0$, $\vf_i(\br)$ is suppressed.  On
defect bonds with $\dis_i(\br)>0$, $\vf_i(\br)$ has finite
fluctuations.  For $\nabla_i \vS (\br) =0$ one would have
\begin{eqnarray}
  \langle f_i^a(\br) f_j^b(\br') \rangle =
  \delta_{ij}^{ab} 
  \delta_{\br,\br'} \frac TJ \dis_i(\br)  .
\label{f.corr}
\end{eqnarray}
However, it is crucial to retain the full correlations between the
fluctuations of $\vS$ and $\vf_i$. The spin-wave saddle-point equation
\begin{eqnarray}
  \nabla_i \vS (\br) = \vf_i(\br) 
  \label{pot.ff}
\end{eqnarray}
immediately shows that $\vf_i$ induces a canting of the spin field.
If $\vf_i(\br)$ is considered as fixed and nonvanishing only on a
single bond, it induces a dipolar spin texture.  It is important to
notice that the original spin rotation symmetry of Hamiltonian
(\ref{H}) is preserved in the transformed model (\ref{H.sf}) only if
$\vf_i$ is rotated simultaneously with $\vS$.

Instead of solving the full problem of two fluctuating fields, Refs.
\onlinecite{Glazman+90,Cherepanov+99} proceed with additional
assumptions about the nature of $\vf_i$.  Glazman and
Ioselevich\cite{Glazman+90} (GI) consider a Hamiltonian, where
$\vf_i(\br)$ has a \textit{fixed length} on the defect bonds.  Only
the orientation of $\vf_i(\br)$ is considered as a thermal degree of
freedom. Thereby the rotation symmetry is preserved. However, fixing
the length of $\vf_i(\br)$ contradicts to Eq. (\ref{f.corr}) which
shows that also the magnitude of $\vf_i(\br)$ is a thermally
fluctuating quantity that vanishes in the limit of zero temperature.
This means that in the approach of GI the strength of disorder is
\textit{overestimated} at low temperatures.  This may be a reason for
which GI find a reentrant temperature dependence of $\xi$.

On the other hand, Cherepanov \etal\cite{Cherepanov+99} consider
$\vf_i$ as a \textit{quenched} field with Gaussian correlations.  This
is in contradiction to the annealed nature of $\vf_i$ as is revealed
by the transformed model (\ref{hs-trafo}).  In addition, spin rotation
symmetry is explicitly broken.  This treatment is based on the
assumption that the spin textures freeze at low temperatures.
However, a spontaneous breaking of this symmetry in $d\leq 2$ is ruled
out by the Mermin--Wagner theorem.\cite{Mermin+66} Thus, one may worry
that disorder effects may be \textit{overestimated} also in this
approach.  Possibly, the artificial symmetry breaking is related to
the spurious generation of random fields in a replica treatment as
found by Cherepanov \etal\cite{Cherepanov+99}

\section{RG analysis}
\label{sec.th}

Although the Hubbard--Stratonovich transformation was useful to relate
previous work to the original model, the introduction of $\vf$ is
accompanied with additional difficulties.  Therefore we choose to
continue to work with the original model where the field $\vf$ is
integrated out.  Before we perform a renormalization-group (RG)
analysis, we use the standard replica trick to treat disorder.

\subsection{Replica representation}

After $n$-fold replication the Hamiltonian reads
\begin{subequations}
\begin{eqnarray}
  H^{(n)} =\frac 12 \sum_{\br,i} [1-\dis_i(\br)] J \psi_i(\br) 
\end{eqnarray}
with the abbreviation
\begin{eqnarray}
  \psi_i(\br):= \sum_{\alpha=1}^n [\nabla_i \vS^\alpha (\br)]^2.
\end{eqnarray}
\end{subequations}
(Upper Greek indices label replicas.) Since we assume that the
probability distribution of $\dis_i(\br)$ is uncorrelated and
identical for all bonds, disorder averaging leads to the local and
translation symmetric Hamiltonian ($\cH$ includes the factor $1/T$)
\begin{eqnarray}
  \cH=  \sum_{\br,i} 
  \left\{\frac 12 K \psi_i(\br) - R(\psi_i(\br))\right\}.
  \label{H.rep}
\end{eqnarray}
The cumulant function $R$ is specified by
\begin{eqnarray}
  R(\psi_i(\br)) := \ln  \ \overline { \exp[\dis_i(\br) K \psi_i(\br) / 2]}
\end{eqnarray}
and depends implicitly on $K:=J/T$.  For the special case of bimodal
disorder (\ref{bimodal})
\begin{subequations}
\begin{eqnarray}
  R(\psi) &=& \ln \left[ 1-p+p \ e^{\dis K \psi/2} \right] 
  \\
  &\approx&
  \left\{
    \begin{array}{ll}
      \frac 12 p \dis K \psi & \textrm{ for } \dis K \psi \to 0 ,
      \\
      \frac 12 \dis K \psi + \ln p & \textrm{ for } \dis K \psi \to
      \infty .
    \end{array}
  \right.
\end{eqnarray}
\label{R.bim}
\end{subequations}
Note that naturally $R(\psi)=0$ in the absence of disorder (for $p=0$
or $\dis=0$) and that $R(\psi)=\frac 12 \dis K \psi$ for $p=1$, which
amounts to an unfrustrated dual model with a corresponding stiffness
$(1- \dis ) J$.  In the general case, $R(\psi)$ is a nonlinear
function which has a linear asymptotics for small and large arguments,
cf. Rq. (\ref{R.bim}).  

Remarkably, the energy depends on $\psi$ only
through the combination
\begin{eqnarray}
  h(\psi):=K\psi-2 R(\psi).
\end{eqnarray}
\textit{A priori}, it is not clear whether spin fluctuations are
governed by the behavior of $h(\psi)$ at large $\psi$ or at small
$\psi$.

The stability of the ferromagnetic state with respect to large-scale
spin-wave deformations depends on $h(\psi)$ at small $\psi$ and
requires $h'(0)>0$, i.e., $ p \dis < 1$.  (For $ p \dis > 1$ the dual
antiferromagnetic ground state is locally stable.)  This condition is
equivalent to the requirement that the average exchange coupling
(\ref{J.av}) should remain ferromagnetic.

While the effective stiffness is defined from the slope of the
function $h(\psi)$ at $\psi=0$, its behavior $h(\psi) \sim (1-\dis)K
\psi$ for $\psi \to \infty$ (which is equivalent to $T \to 0$)
reflects the presence of frustration. For $\dis<1$ --- i.e., in the
\textit{absence} of frustration --- $h(\psi)$ is positive for all
$\psi$ and fluctuations can renormalize the stiffness to a smaller but
positive value.  For $\dis>1$ --- i.e., in the \textit{presence} of
frustration --- $h(\psi)$ is negative at large $\psi$ and fluctuations
can renormalize the stiffness to zero, signaling the destruction of
magnetic order.

In the replica representation, the canting instability for a single
defect can be retrieved easily.  To this end, one has to keep the
cumulant function (\ref{R.bim}) only on a single bond and to switch it
off on all other bonds.  In the limit $T\to 0$ (considering the
statistical weight for fixed $n$ and an fixed but arbitrary spin
configuration), the replicas decouple since $R(\psi)\approx \frac 12
\dis K \psi + \ln p$. The exchange coupling of the defect bond within
each replica is $(1-\dis)J$ independent of $p$. Thus, canting occurs
in the case $p<1$ for $\dis>\dis_\textrm{single}$ as in the case
$p=1$. While the threshold is independent of $p$, the
disorder-averaged dipole moment depends on $p$.  Since $\psi$ becomes
arbitrarily small slightly above the canting threshold, this dipole
moment cannot be calculated from the large-$\psi$ limit of $h$.

Therefore it is in general important to retain the \textit{global}
functional form of $h(\psi)$ [or, equivalently, $R(\psi)$].  An
approximation of this function by a Taylor series near $\psi=0$ would
amount to an expansion in cumulants of the defect distribution.  The
relevance of $R(\psi)$ for large $\psi$ shows that one could miss
essential physics by dropping high-order cumulants and -- in
particular -- by using a Gaussian distribution for $\dis$.  Even
worse: the function $R(\psi)$ is \textit{nonanalytic}, i.e., its
behavior at large $\psi$ is outside the radius of convergence of a
cumulant expansion.

\subsection{Flow equations}

In order to address the question of magnetic order in the presence of
general fluctuations of $\psi$, we generalize the
renormalization-group analysis of Polyakov\cite{Polyakov75} to the
replicated model.

Instead of working with the function $h(\psi)$, it is more physical to
use an effective bare stiffness and an effective cumulant function
defined by
\begin{subequations}
\label{split}
\begin{eqnarray}
  K_0 &:=&K-2 R'(0) = (1-p \dis)K,
  \\
  R_0(\psi)&:=&R(\psi)-\psi R'(0) = R(\psi)- \frac 12 p \dis K \psi 
\end{eqnarray}
\end{subequations}
as to remove the linear contribution of $R$, which represents a
trivial renormalization of the stiffness. The replacement of the
original quantities by the effective ones leaves the energy
(\ref{H.rep}) invariant since $K\psi-2 R(\psi) =K_0\psi-2 R_0(\psi) $.

Following Polyakov's analysis of the pure system,\cite{Polyakov75} we
renormalize the model by a momentum-shell integration.  So far, the
Hamiltonian (\ref{H.rep}) was written in a way explicitly retaining
the lattice of the spin sites.  We choose the lattice spacing as unit
of length and go over to a continuum representation by smoothly
interpolating the field $\vS(\br)$ between the lattice sites
(preserving the normalization $\vS^2(\br)=1$ everywhere).  In this
limit, partial spatial derivatives replace the differences in
$\psi_i(\br)= \sum_{\alpha=1}^n [\partial_i \vS^\alpha (\br)]^2$ and
\begin{eqnarray}
  \cH =  \int \di^dr  \sum_i 
  \left\{\frac 12 K \psi_i(\br) - R(\psi_i(\br))\right\}.
  \label{H.rep.cont}
\end{eqnarray}
The replacement of differences by partial derivatives and of the sum
by an integral should be a reasonably good approximation, i.e., the
replacement of $\psi$ should lead to relative errors small compared to
unity for an \textit{arbitrary} spin configuration (even with
inhomogeneities on the scale of the lattice spacing). A small relative
error in $\psi$ is equivalent to a small relative misrepresentation of
$J$ and/or $\dis$ since these quantities enter the Hamiltonian only as
a product with $\psi$.  We further use the approximate replacement of
the cubic Brillouin zone by a spherical one.  Thus, to preserve the
volume of the Brillouin zone, its radius $\Lambda$ has to be fixed by
$\Lambda^d = d (2 \pi)^d/\cS_d$ with $\cS_d:= 2
\pi^{(d/2)}/\Gamma(d/2)$ the surface of the $d$ dimensional unit
sphere.  In $d=2$ specifically, $\Lambda^2=4\pi$.  In order to
demonstrate that the fundamental frustration mechanism does not get
lost due to the continuum approximation, we briefly rederive the
canting threshold in Appendix \ref{app.cont}.

According to the scheme of the momentum-shell renormalization group,
we now integrate out the spin modes with wave vectors in the shell
$\Lambda e^{-\di \ell} < k \leq \Lambda$. Thereby the original spin
field $\vS^\alpha$ is mapped onto the slowly varying background field
$\vs^\alpha$.  They are related by
\begin{eqnarray}
  \vS^\alpha = \sqrt{1-\chi_a^\alpha \chi_a^\alpha} \ \vs^\alpha + 
  \chi_a^\alpha \ve_a^\alpha
\end{eqnarray}
with the vector fields $\ve_a^\alpha$ forming a local orthonormal
basis $\{\ve_1^\alpha, \cdots, \ve_{N-1}^\alpha, \vs^\alpha\}$ at each
site $\br$ in each replica $\alpha$.  We employ the sum convention for
pairs of Latin indices ($a, b =1,\cdots,N-1$) only.  The field
$\chi_a^\alpha$ generates an infinitesimal spin rotation and has
contributions only from wave vectors in the momentum shell.

Derivatives of basis vectors can be expanded in the local bases
\begin{subequations}
\begin{eqnarray}
  \partial_i \vs^\alpha &=&B_{ia}^\alpha \ve_a^\alpha
  \\
  \partial_i \ve_a^\alpha &=& - B_{ia}^\alpha \vs^\alpha + 
  A_{iab}^\alpha \ve_b^\alpha 
\end{eqnarray}
\end{subequations}
in terms of potentials $A$ and $B$.  The arbitrariness of the choice
of the vectors $\ve_a$ is reflected in a gauge invariance of the
potentials; the gauge transformations are local rotations around
$\vs$.\cite{Polyakov87} One can exploit this gauge symmetry to show
that the potential $A$ corresponds to higher order derivatives of the
spin field.\cite{Polyakov87,Cherepanov+99} We ignore such
contributions and therefore omit the potential $A$ from now on.

For the further calculations, $\chi$ can be considered as small since
we consider an infinitesimal momentum shell. In addition, we treat
temperature and disorder, which drive the fluctuations of $\chi$, as
small.  In order to expand the Hamiltonian in $\chi$, we first
consider
\begin{eqnarray}
  \partial_i \vS^\alpha &=& B_{ia}^\alpha \ve_a^\alpha 
  \nonumber   \\
  &&-\chi_a^\alpha B_{ia}^\alpha \vs^\alpha 
  +\partial_i \chi_a^\alpha \ve_a^\alpha
  \nonumber     \\
  &&-\chi_a^\alpha \partial_i \chi_a^\alpha \vs^\alpha 
  - \frac 12 \chi_b^\alpha \chi_b^\alpha
  B_{ia}^\alpha \ve_a^\alpha + O(\chi^4)
\end{eqnarray}
and, ordered by powers of $\chi$,
\begin{subequations}
\begin{eqnarray}
  \psi_i&=&\psi_i^{(0)}+\psi_i^{(1)}+\psi_i^{(2')}+\psi_i^{(2'')}+O(\chi^3),
  \\
  \psi_i^\0 &=& \sum_\alpha B_{ia}^\alpha B_{ia}^\alpha 
  = \sum_\alpha (\partial_i \vs^\alpha)^2,
  \\
  \psi_i^\1 &=& \sum_\alpha  2 \partial_i \chi_a^\alpha  B_{ia}^\alpha ,
  \\
  \psi_i^{(2')} &=& \sum_\alpha \partial_i \chi_a^\alpha
  \partial_i \chi_a^\alpha,
  \\
  \psi_i^{(2'')} &=& \sum_\alpha \bigg\{ 
  \chi_a^\alpha  \chi_b^\alpha B_{ia}^\alpha B_{ib}^\alpha 
  - \chi_b^\alpha \chi_b^\alpha B_{ia}^\alpha B_{ia}^\alpha
  \bigg\}.
 \end{eqnarray}
\end{subequations}
An expansion of the energy density for small $\chi$ gives
\begin{eqnarray}
  h(\psi_i) &=& h(\psi_i^\0) + h'(\psi_i^\0) [ \psi_i^\1 +
  \psi_i^{(2')} + \psi_i^{(2'')}        ] 
  \nonumber \\ &&
  + \frac 12  h''(\psi_i^\0) [ \psi_i^\1]^2 + O(\chi^3).
\end{eqnarray}
Substituting this expression back into (\ref{H.rep.cont}), we rewrite
\begin{subequations}
\begin{eqnarray}
  \cH &=& \cHf + \cHi 
  \\
  \cHf &=& \int \di^dr \sum_i \bigg\{ \frac 12 K_\ell
  [\psi_i^\0 +  \psi_i^{(2')}]
  \nonumber \\ &&
  - R_\ell(\psi_i^\0)
  \\
  \cHi &=& \int \di^dr \sum_i\bigg\{ 
  \frac 12 K_\ell [\psi_i^\1 + \psi_i^{(2'')}]
  \nonumber \\ &&
  - R'_\ell(\psi_i^\0)  [\psi_i^\1 + \psi_i^{(2')}+ \psi_i^{(2'')}]
  \nonumber \\ &&
  - \frac 12 R''_\ell(\psi_i^\0)  [\psi_i^\1]^2
  \bigg\} + O(\chi^3).
\end{eqnarray}
\end{subequations}
Hereby we have separated the ``free'' and ``interaction''
contributions in a way such that $\cHi$ vanishes for $\chi=0$ and that
$\cHf$ contains the bilinear selfinteraction of $\chi$.

Due to the energy contribution $\cHi$ the fluctuations cannot be
integrated out exactly.  In analogy to the treatment of the pure
system, we apply standard perturbation theory to $\cHi$. In principle,
this can be done in a systematic way at low temperatures, where $\chi
\sim T^{1/2}$.  Aiming at the analysis of the stability of magnetic
order for low temperature and weak disorder, we retain only the
renormalization effects to first order in $\cHi$.  Integration over
$\chi$ leads to an infinitesimal renormalization of the Hamiltonian
\begin{eqnarray}
  d \cH= \langle \cHi \rangle + O(\cHi^2)
  \label{H.av}
\end{eqnarray}
where the average is over the fluctuations of $\chi$ weighted by
$\cHf$ only.

Using the averages
\begin{subequations}
\begin{eqnarray}
  \langle \partial_i \chi_a^\alpha \partial_j \chi_b^\beta \rangle
  &=& \frac 1 {K_\ell} \delta_{ij} \delta_{ab} \delta^{\alpha \beta} \di \ell,
  \\
  \langle \chi_a^\alpha \chi_b^\beta \rangle
  &=&
  \frac d {K_\ell \Lambda^2}  \delta_{ab} \delta^{\alpha \beta} \di \ell,
\end{eqnarray}
\end{subequations}
one finds
\begin{subequations}
\begin{eqnarray}
  \langle \psi_i^\1 \rangle &=& 0,
  \\
  \langle \psi_i^{(2')} \rangle &=& O(n) ,
  \\
  \langle \psi_i^{(2'')} \rangle &=& 
  -(N-2) \frac d{K_\ell \Lambda^2} \psi_i^\0 \di \ell,
  \\
  \langle [\psi_i^\1]^2 \rangle &=& \frac 4{K_\ell} \psi_i^\0 \di \ell.
\end{eqnarray}
\end{subequations}
Separating the flow of $K_\ell$ and of $R_\ell$ by the requirement
$R_\ell'(0)=0$ as in Eqs. (\ref{split}) for the bare quantities, we
finally obtain the flow equations
\begin{subequations}
\begin{eqnarray}
  \frac \di{\di\ell} K_\ell&=& (d-2) K_\ell - \frac {N-2}{\Lambda^2/d} 
    - \frac  4{K_\ell} R''_\ell(0), 
  \\
  \frac \di{\di\ell} R_\ell(\psi)&=& d R_\ell(\psi) 
  - (2 + \frac{N-2}{K_\ell \Lambda^2/d}) \psi R'_\ell(\psi) 
  \nonumber \\ &&
  + \frac 2{K_\ell} \psi [R''_\ell(\psi)-R''_\ell(0)].
\end{eqnarray}
\end{subequations}
Terms explicitly proportional to the number of replicas $n$ have been
dropped.  A rescaling $\br \to e^{\di \ell} \br$ of lengths has been
included in order to keep the value of the cutoff fixed.

For arbitrary temperature, the flow of $K$ can be interpreted as a
flow of the spin stiffness at fixed temperature.  Using the
dimensionless stiffness $j_\ell:=K_\ell/K$, the dimensionless
temperature $t:=T/J=1/K$, the rescaled field $\phi:=\dis K \psi$ (we
recall that $K$ and $\dis$ are unrenormalized quantities), and
$\hR_\ell(\phi):=R_\ell(\psi)$, we rewrite the flow equations as
\begin{subequations}
\label{flow}
\begin{eqnarray}
  \frac \di{\di\ell} j_\ell&=& (d-2) j_\ell - \frac{N-2}{\Lambda^2/d}t - 
  \frac{4 \dis^2}{j_\ell} \hR''_\ell(0),
  \label{flow.j}
  \\
  \frac \di{\di\ell} \hR_\ell(\phi)&=& d \hR_\ell(\phi) 
  - (2 + \frac{N-2}{\Lambda^2/d} \frac t {j_\ell}) \phi \hR'_\ell(\phi) 
  \nonumber \\ &&
  + \frac{2 \dis}{j_\ell} \phi [\hR''_\ell(\phi)-\hR''_\ell(0)].
  \label{flow.hR}
\end{eqnarray}
\end{subequations}
In this form, temperature and disorder strength $\dis$ appear as
explicit parameters.  The flow equations (\ref{flow}) have to be
solved with the initial conditions
\begin{subequations}
\begin{eqnarray}
   j_0&=&1-p \dis ,
   \\
   \hR_0(\phi)&=& \ln \left[ 1-p+p \ e^{\phi/2} \right] - \frac 12 p
   \phi .
\end{eqnarray}
\end{subequations}

Corresponding to the neglect of higher orders of $\cHi$ in Eq.
(\ref{H.av}), these flow equations contain renormalization effects
only to the leading order in temperature and disorder.  Higher orders
of perturbation theory would certainly generate higher order
contributions as well as a more complicated functional form of the
Hamiltonian.

Anticipating that $\hR_\ell''(0)\geq 0$ is preserved under the flow,
Eq. (\ref{flow.j}) shows that both thermal fluctuations and disorder
tend to reduce the effective stiffness. 

\begin{figure}
  \includegraphics[width=0.9\linewidth]{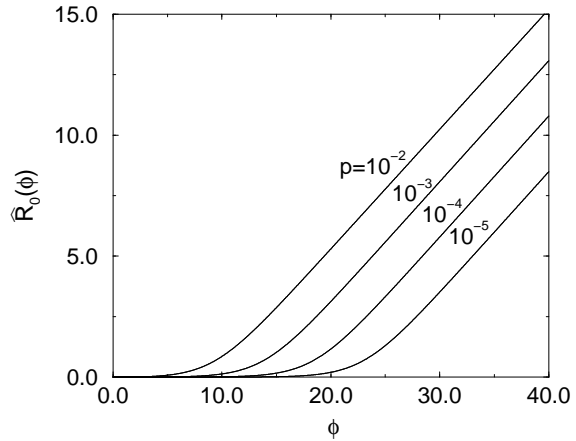}
  \caption{Plot of the initial function $\hR_0(\phi)$ for the values
    $p=10^{-2}$, $10^{-3}$, $10^{-4}$, and $10^{-5}$.  Note that this
    nonanalytic function is very small up to $\phi^*=2 \ln
    \frac{1-p}p$, where it assumes slope that is approximately
    independent of small $p$.}
  \label{fig.hR}
\end{figure}

In the given order, several features of the flow equations are
remarkable.  The initial function $\hR_0(\phi)$ depends only on $p$
(cf. Fig \ref{fig.hR}). The crossover from the linear regime at small
$\phi$ to the linear regime at large $\phi$ occurs at $\phi^*=2 \ln
\frac{1-p}p$, where the curvature $\hR_0''(\phi) = 1/16
\cosh^2[(\phi-\phi^*)/4]$ has its maximum $\hR_0''(\phi^*) = \frac
1{16}$. The flow equation of the stiffness couples to disorder only
through $\hR_0''(0)$.  For $\dis=0$, this coupling vanishes, while
$\hR_\ell(\phi)=0$ for $p=0$.  The flow equations depend on $N$ only
at finite temperatures (unlike the flow equations of Ref.
\onlinecite{Cherepanov+99}).  Thus, the properties at zero temperature
are expected to be \textit{independent} of $N$.  However, one has to
keep in mind that the equations apply only to spins with a continuous
rotation symmetry ($N \geq 2$) and that the renormalization scheme
ignores the effects of topological defects, which are known to be
particularly important for $N=2$.




\section{Results and discussion}
\label{sec.res}

In order to determine the large-scale properties of the model, we
numerically integrate the coupled flow equations (\ref{flow}) for
$j_\ell$ and $\hR_\ell(\phi)$ from $\ell=0$ to large $\ell$.  If
$j_\ell$ converges for $\ell \to \infty$ to a finite value much larger
than $t$, the system is in an ordered state with finite renormalized
stiffness and infinite correlation length, i.e.  $\overline{ \langle
  \vS(\br) \vS(\br') \rangle}$ decays slower than exponentially with
the distance $|\br-\br'|$.  If, on the other hand, $j_\ell$ becomes of
order $t$ on a finite scale $\ell=\ell^*$, the system is disordered
and we identify the correlation length as
\begin{eqnarray}
  \xi = A \ e^{\ell^*}
\label{xi}
\end{eqnarray}
with some constant $A$ of the order of the lattice spacing.  Following
previous references,\cite{Chakravarty+89,Cherepanov+99}  we
specifically define this scale from
\begin{eqnarray}
  j_{\ell^*}= \frac t{2\pi}.
\end{eqnarray}

The critical disorder strength $\dis_\textrm{FM}$ separating a
ferromagnetically ordered phase from a disordered phase can be
identified with the line where $\xi$ diverges.  Unfortunately, this
criterion does not allow for a distinction between true long-range
order and quasi-long-range order in the ferromagnetic phase.

\subsection{Zero temperature}

\begin{figure}
  \includegraphics[width=0.9\linewidth]{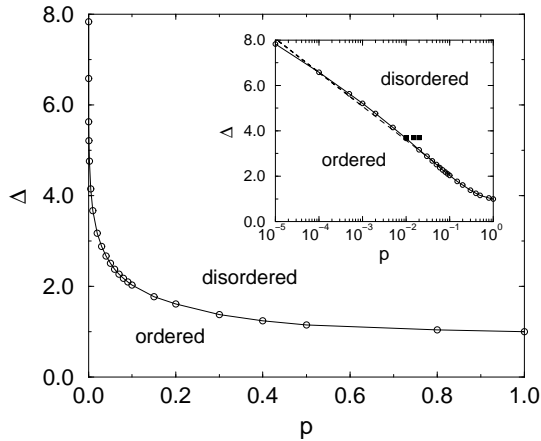}
  \caption{Two-dimensional phase diagram at $T=0$ in a linear  and
    semi-logarithmic plot.  The solid line represents the transition
    line $\dis_\textrm{FM}(p)$ from the magnetically ordered to the
    disordered phase.  In the inset the dashed line represents the
    linear fit (\ref{law.log}) and the three filled squares visualize
    the points with $\dis=3.7$ with $p=0.01$, $0.015$ and $0.02$ used
    in Sec. \ref{sec.compare} for comparison with experiments.}
  \label{fig.zero}
\end{figure}

At $T=0$, we find the transition line $\dis_\textrm{FM}(p)$ as shown in
Fig. \ref{fig.zero}.  
This transition has the following features:
  
\textit{Regime of small $p$.} For dilute disorder, $p \to 0$, we find
a slow but unbounded increase of $\dis_\textrm{FM}(p)$.  In the range
$10^{-5} \leq p \leq 0.1$ the line follows roughly the relation [cf.
the dashed line in the inset of Fig.  \ref{fig.zero}]
\begin{eqnarray}
  \dis_\textrm{FM}(p) \approx  0.606 + 0.648 \ln \frac 1p .
  \label{law.log}
\end{eqnarray}
However, in the limit $p \to 0$, $\dis_\textrm{FM}(p)$ appears to
increase slower than logarithmically.

This finding implies order below a finite defect concentration even
for arbitrarily strong defect bonds.  This behavior is in disagreement
with Refs.  \onlinecite{Glazman+90} and \onlinecite{Cherepanov+99},
where disorder (with $\dis > \dis_\textrm{single}$) was suggested to
destroy order for infinitesimally small $p$.  As already stated in
Sec.  \ref{sec.prev}, we believe that the effects of disorder are
overestimated in both previous approaches because of special
assumptions about the nature of the bivectorial field $\vf_i$.

For the special case $N=2$, where this disagreement persists, further
references can be included in the comparison.  (We assume here that XY
and Heisenberg systems should have a similar phase diagram at $T=0$
since our lowest-order flow equations are independent of $N$.
Topological defects are ignored in our work as well as in Ref.
\onlinecite{Cherepanov+99}.  Their presence may further reduce $\xi$.)
The way how disorder is treated by Cherepanov
\etal\cite{Cherepanov+99} implies that their analysis actually
describes an XY model with random phase shifts.  Various recent work
(see e.g. \onlinecite{Nattermann+95}) has provided evidence that
quasi-long-range order should exist for weak disorder (even in the
presence of vortices).  This observation is consistent with our flow
equations but contradicting the flow equations of Cherepanov
\etal\cite{Cherepanov+99} From their numerical data, Gawiec and
Grempel\cite{Gawiec+91} argue for a disordered phase for $\dis >
\dis_\textrm{single}$ and $p>0$, i.e., for $\dis_\textrm{FM}(0) = 2$.
While this conclusion is again in disagreement with our result, the
numerical data are not: Gawiec and Grempel\cite{Gawiec+91} present
data for $\dis=4$ as the only value with $\dis>2$ and they demonstrate
the absence of order only for $p \gtrsim 0.02$.  For $\dis=4$, we find
order at very small $p \lesssim 0.0062$, which actually is
\textit{not} excluded by the numerical data (cf. Fig. 17 of Ref.
\onlinecite{Gawiec+91}).

\textit{Regime of large $p$.}  For $p \to 1^-$ we find that
$\dis_\textrm{FM}(p) \to 1^+$ in a smooth way.  While this is
qualitatively correct, a horizontal segment with
$\dis_\textrm{FM}(p)=1$ beyond the percolation transition at $p=\frac
12$ is absent.  However, in this range we find a $\dis_\textrm{FM}(p)$
which is only slightly larger, i.e., this is a quantitative effect
which may be attributed to the continuum approximation as argued below
Eq. (\ref{H.rep.cont}).  In addition, the regime near $\dis=1$ and $p
\gtrsim \frac 12 $ is a regime of strong disorder in the sense of Sec.
\ref{sec.finite} since $\sigma^2 \gtrsim 1$.  There our flow equations
are not quantitatively reliable also because of the lowest-order
truncation.  On the other hand, the transition at small $p$ should be
well described since Eq.  (\ref{est.FM}) is satisfied, $p
\dis^2_\textrm{FM}(p) \to 0 $ for $p \to 0$.

\begin{figure}
  \includegraphics[width=0.9\linewidth]{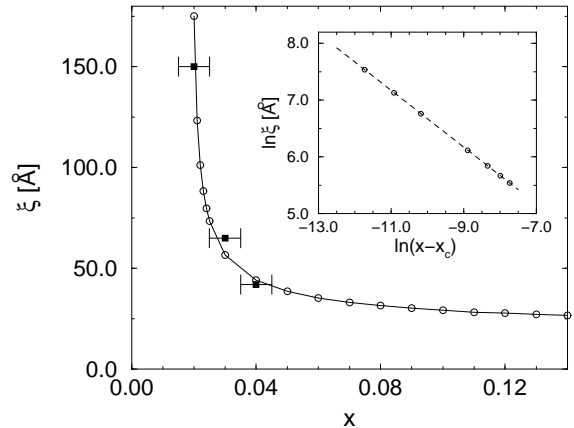}
  \caption{
    Plot of $\xi(x)$ for $T=0$ with $\dis=3.7$ and $A=22${\AA} in
    order to fit the experimental data (filled squares with error
    bars) of Ref.  \onlinecite{Keimer+92b}.  The inset shows a
    double-logarithmic plot of $\xi$ as a function of $x-x_\textrm{c}$
    for $\dis=3.7$ with $x_\textrm{c}=0.019062$.  Open circles
    connected by a line represent $\xi$ calculated numerically from
    the flow equations. The dashed line is the best linear fit with
    slope $-\nu=-0.50007$.}
  \label{fig.xi_x}
\end{figure}

When the transition line is approached at finite $p$, the correlation
length displays a divergence
\begin{eqnarray}
  \xi(\dis,p)  \sim [\dis - \dis_\textrm{FM}(p)]^{-\nu} .
\label{div.xi}
\end{eqnarray}
From the numerical integration of our flow equations we find with a
mean-field like exponent $\nu = 0.500(1)$ [cf. the inset of Fig.
\ref{fig.xi_x}].

\subsection{Finite temperatures}

Temperature enters the flow equations in two places where it could
lead to contrary effects. In Eq.  (\ref{flow.j}), an increase of
temperature leads to a faster renormalization of the spin stiffness to
smaller values [ignoring the temperature dependence of
$\hR''_\ell(0)$].  On the other hand, in Eq.  (\ref{flow.hR})
temperature tends to suppress $\hR$, which might in turn reduce the
efficiency of disorder in suppressing the spin stiffness.  However,
from our flow equations, we always find that thermal fluctuations
reduce stiffness and therefore also $\xi$.

More precisely, $\xi$ decreases monotonously with increasing
temperature.  Thus, a reentrant temperature dependence as found by
GI\cite{Glazman+90} is absent in the present treatment.  We
attribute this discrepancy to the fact that GI keep the length of the
bivector $\vf_i$ fixed, whereas its typical length should vanish at
low $T$ according to Eq. (\ref{f.corr}).  Thereby, with decreasing
temperature, the strength of disorder is increasingly overestimated,
giving way to an apparent reentrance.

\subsection{Comparison with experiments}
\label{sec.compare}

We now turn to check the consistency of our theory with measurements
on La$_{2-x}$Sr$_x$CuO$_4$.  To allow for a comparison of our results
with those of Cherepanov \etal\cite{Cherepanov+99} we refer to the
same experimental data by Keimer \etal\cite{Keimer+92b} for $x=0.02$,
$0.03$ and $0.04$.  For the moment we assume that interlayer couplings
and spin anisotropies can be neglected and come back to this issue
later on.

The comparison of our results for $\xi$ with experiments involves four
parameters, $p$, $\dis$, $J$, and $A$.  As already stated above, our
model parameter $p=x/2$ is directly related to the dopant
concentration $x$.  $A$ is a length scale of the order of the lattice
spacing.  In principle, this parameter can depend on temperature and
on disorder itself (compare the discussion in Ref.
\onlinecite{Cherepanov+99}).  Such dependences could modify the
function $\xi(x,T)$ in a subdominant way and would involve 
additional assumptions and parameters.  We refrain from including such
dependences for the purpose of the subsequent semiquantitative
comparison.

At $T=0$, $J$ does not enter the flow equations since it simply sets
the energy scale and $\dis$ is the only unknown model parameter which
enters the numerical calculation of $\ell^*$.  From the consideration
of the superexchange across a defect bond one expects $\dis \gg
1$,\cite{Aharony+88} i.e., a value clearly above the canting
instability.  The finiteness of the measured values of $\xi$ for $x
\geq 0.02$ implies that the data points lie in the disordered phase,
i.e., $\dis > \dis_\textrm{FM}(p=0.01)\approx 3.67$.

We have determined values of the parameters $\dis$, $A$, and $J$ from
the requirement that our theoretical values for $\xi$ should be
consistent with experimental data.  In view of the given experimental
errors and the approximate nature of our RG calculation, we found a
satisfactory agreement for
\begin{subequations}
  \label{param}
  \begin{eqnarray}
    \dis &=& 3.7,
    \\
    J &=& 240 \textrm{ K},
    \\
    A &=& 5.8 a,
  \end{eqnarray}
\end{subequations}
where $a=3.8${\AA} is the lattice spacing.  In Fig. \ref{fig.xi_x} we
compare our theory with data for $\xi$ as a function of $x$ at $T=0$.
The inset shows shows for $\dis=3.7$ a double-logarithmic plot of
$\xi(x)$ which reveals the mean-field like divergence of $\xi$
according to Eq. (\ref{div.xi}) near the order-disorder transition.

\begin{figure}
  \includegraphics[width=0.9\linewidth]{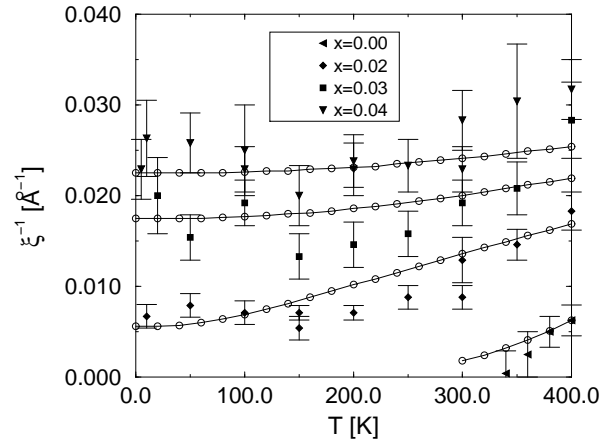}
  \caption{
    Plot of $\xi^{-1}(T)$ for $x=0$, $0.02$, $0.03$, and $0.04$.
    Symbols with error bars are experimental data from Ref.
    \onlinecite{Keimer+92b}.  Open circles connected by lines
    represent our theory.}
  \label{fig.xi.T}
\end{figure}

The temperature dependence of $\xi$ is compared in Fig.
\ref{fig.xi.T}.  Keeping in mind the strong fluctuations of the
experimental data, they can be considered as consistent with our
analysis.  However, the theoretical dependence of $\xi$ on $T$ and $p$
does not quantitatively confirm the empirical
formula\cite{Keimer+92b,Shirane+94}
\begin{eqnarray}
  \xi^{-1}(T,p)=\xi^{-1}(T,0)+\xi^{-1}(0,p),
\end{eqnarray}
which would imply that in Fig. \ref{fig.xi.T} the curves for different
$p$ should differ only by a vertical shift.  In contrast, we find that
thermal fluctuations lead to a stronger increase of $\xi^{-1}$ for
smaller $p$.

\subsection{Coupled layers}

We now address the effects of a very weak interlayer coupling
$J^\perp$.  For La$_{2}$CuO$_4$, a ratio $J^\perp/J = 5 \times
10^{-5}$ was determined from experiments.\cite{Keimer+92a} From a
simple scaling analysis,\cite{Kosterlitz+78} one immediately obtains
the flow equation
\begin{eqnarray}
  \frac \di{\di\ell} J^\perp_\ell = 2 J^\perp_\ell 
\end{eqnarray}
which shows the strong relevance of this coupling. From the condition
that the interlayer coupling becomes comparable to the intralayer
coupling, $J^\perp_\ell=J$, one can fix a dimensional crossover scale
\begin{eqnarray}
  \ellcr= \frac 12 \ln \frac{J}{J^\perp}.
\end{eqnarray}

\begin{figure}
  \includegraphics[width=0.9\linewidth]{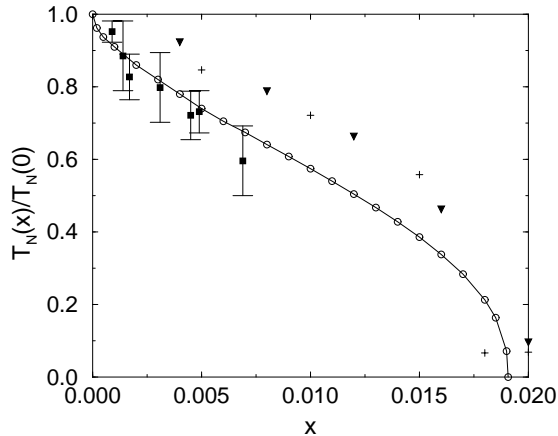}
  \caption{
    Plot of our numerical result for $T_\textrm{N}(x)$ (open circles
    connected by a line).  For comparison, squares,\cite{Chen+91}
    triangles,\cite{Cho+93} and crosses\cite{Saylor+90} represent
    experimental data (cf. the main text).}
  \label{fig.neel}
\end{figure}

Comparing this scale with the correlation length (\ref{xi}) obtained
in the \textit{absence} of the layer coupling, one expects that the
coupling actually is irrelevant for $\ell^*<\ellcr$, where 2D
fluctuations on small scales renormalize $J^\perp$ to zero.  On the
other hand, for $\ell^*>\ellcr$ fluctuations become three-dimensional
on scales larger than $\ellcr$. Per definition of $\ell^*$, the
exchange coupling on the scale $\ellcr$ is still large compared to
temperature such that magnetic long-range order should be stable.
Therefore, the location of three-dimensional ordering in the $(x,T)$
plane can be determined from the implicit condition
\begin{eqnarray}
  \ell^*(x,T)= \frac 12 \ln \frac{J}{J^\perp}.
\end{eqnarray}
We have evaluated this condition numerically for $J^\perp/J = 5 \times
10^{-5}$ and the parameter set (\ref{param}).  The resulting N\'eel
temperature is plotted in Fig.  \ref{fig.neel} as a function of
doping.  The transition temperature is normalized by its value
$T_\textrm{N}(0)\approx 300$K in the absence of disorder, which
essentially reflects the value of $J$.  While the precise value of $J$
may vary to some extent with the employed fitting procedure for the
parameter set, the shape of normalized transition line is very robust.
At $T=0$, the critical disorder strength $x_\textrm{3D}$ is increased
by the interlayer coupling only slightly over $x_\textrm{2D}\approx
0.1906$, $(x_\textrm{3D}- x_\textrm{2D})/ x_\textrm{2D} \approx
10^{-2}$, because of the extremely slow divergence of $\xi$ for $x \to
x_\textrm{2D}^+$.

In comparison to the theory in Ref. \onlinecite{Cherepanov+99}, we
find that $T_\textrm{N}(x)$ decays less abruptly near $x_\textrm{3D}$.
We oppose our results to data from Hall measurements\cite{Chen+91}
(squares in Fig. \ref{fig.neel}), susceptibility
measurements\cite{Cho+93} (triangles), and PAC
measurements\cite{Saylor+90} (crosses).  The overall agreement is
satisfying, although the experimental data partially suggest that
disorder is less effective in reducing the transition temperature.
This tendency may be attributed to the fact that the samples were
partially oxygen-doped (at higher temperature, oxygen is not quenched
and less effective in generating spin frustration) as well as the fact
that we have neglected the easy-plane spin anisotropy, which also
tends to stabilize the magnetic order.

\section{Conclusions}
\label{sec.concl}

In this article, we have reexamined a classical model for
$N$-component spins ($N\geq 2$) with random exchange couplings.
Thereby we have chosen an approach complementary to previous studies
by Glazman and Ioselevich\cite{Glazman+90} and by Cherepanov
\etal\cite{Cherepanov+99} Special care has been taken to preserve the
quenched nature of disorder and the global spin rotation symmetry.

Our analysis involves approximations which we briefly summarize.  (i)
The originally discrete spin system is represented in a continuum
formulation. Thus, features related to the specific lattice structure
-- such as the location of the percolation transition for $\dis=1$ --
cannot be captured quantitatively. (ii) As in the pure case, the
renormalization scheme accounts only for interactions between spin
waves, ignoring the role of topological defects.  Thus, the degree of
magnetic order may be overestimated (recall that for $N=2$ the pure
system erroneously appears to be ordered at all temperatures if
vortices are neglected). (iii) The flow equations are truncated to
lowest order in temperature and disorder.  In principle, the analysis
could be extended to higher orders. In practice, this extension is
hampered by a much more complicated functional form of the Hamiltonian
which is generated during the flow.

Due to the nonanalytic nature of the cumulant function $R$, we found
it necessary to develop a \textit{functional} renormalization group,
i.e., to keep track of \textit{arbitrarily high cumulants} of the
disorder distribution.  To the best of our knowledge, functional flow
equations for disordered spin systems have been considered previously
only for different types of disorder, in particular for random fields
and random anisotropy.\cite{Fisher85,Feldman99,Feldman00,Feldman01}

The flow equations (\ref{flow}) are the central result of the analytic
part of this work.  For comparison, the analysis of Cherepanov
\etal,\cite{Cherepanov+99} which employs approximations corresponding
to the ones listed above, is restricted to a single disorder parameter
that corresponds to the lowest cumulant of the disorder distribution.

As main physical result from our RG analysis, we find that the
two-dimensional spin system can be magnetically ordered at $T=0$ in
the presence of a sufficiently small but \textit{finite} concentration
of arbitrarily strong defects ($\dis>\dis_\textrm{single}$).  At this
place it is worthwhile to recall that we have identified magnetic
order from the length scale where the spin stiffness is renormalized
down to the scale of thermal fluctuations.  In the absence of explicit
calculations for the spin-spin correlation function it is natural to
assume that this scale coincides with the magnetic correlation length
$\xi$.  Such a calculation would be desirable in order to clarify
whether the ordered phase (with $\xi=\infty$) has quasi-long-range
order or true long-range order.  We have determined $\xi$ by a
numerical integration of our flow equations for the strictly
two-dimensional system as well as for weakly coupled layers.  In the
first case our zero-temperature phase diagram is consistent with
numerical simulations.\cite{Gawiec+91} In the second case, the
calculated dependence of $\xi$ on temperature and on disorder strength
is in good agreement with measurements on cuprates.  In both cases the
comparison was restricted to \textit{finite} length scales given by
the computationally manageable system sizes or by the experimental
error bars, respectively.

Nevertheless, concerning the question of whether magnetic order is
stable against bond disorder in two dimensions on largest scales, our
positive answer, shared by Ref. \onlinecite{Rodriguez+95}, disagrees
with previous negative ones.\cite{Glazman+90,Cherepanov+99} In the end
of Sec. \ref{sec.prev} we have given specific reasons why in our
opinion the previous answers cannot be considered as final.  In view
of the complementary approaches and approximations involved, this
question has to be considered as an open one that calls for additional
future research.

\begin{acknowledgments}
  We gratefully acknowledge discussions with A. Aharony, D. R.
  Grempel, T.  Nattermann, H. Rieger, and B. Rosenow.
\end{acknowledgments}


\appendix

\section{Duality}
\label{app.dual}

For the bimodal distribution (\ref{bimodal}) with $\dis>1$, our system
consists of a fraction $p$ of ``defective'' antiferromagnetic bonds
and a fraction $1-p$ of ``regular'' ferromagnetic bonds. The relative
strength of bonds is $|J_\textrm{AFM}/J_\textrm{FM}|=1-\dis$.  From a
dual point of view, one may say that the system consists of a fraction
$\tp:=1-p$ of ``defective'' ferromagnetic bonds and a fraction
$1-\tp=p$ of ``regular'' antiferromagnetic bonds.  Since for classical
spins thermodynamic properties are invariant under flipping one
sublattice and reversing the sign of the exchange coupling
($J_\textrm{AFM}=:- \tJ_\textrm{FM}$ and $J_\textrm{FM}=:-
\tJ_\textrm{AFM}$),\cite{Fradkin+78} the system is equivalent to a
system with a fraction $\tp:=1-p$ of ``defective'' antiferromagnetic
bonds and a fraction $1-\tp=p$ of ``regular'' ferromagnetic bonds of
relative strength $|\tJ_\textrm{AFM}/\tJ_\textrm{FM}|=1/(1-\dis)=:1-
\tdis$.\cite{Gawiec+93} Thus duality provides a mapping
\begin{subequations}
\begin{eqnarray}
  J &\to& \tJ=(1-\Delta)J,
  \\
  p &\to& \tp=1-p,
  \\
  \dis &\to& \tdis=\frac{\dis}{\dis-1},
\end{eqnarray}
\end{subequations}
for all temperatures.

\section{Continuum limit}
\label{app.cont}

We show that spin frustration mechanism is well captured in the
continuum representation of our model.  To this end, we rederive the
canting instability threshold $\Delta_{\mathrm{single}}=d$ for a
single defect bond. For a single defect bond located at $\mathbf{r}=0$
and oriented in direction $\mathbf{e}_j$ the Hamiltonian reads
\begin{equation}
  H=\frac{J}{2}\int\mathrm{d}^dr\sum_i
  (1-\delta_{ij}\delta(\mathbf{r})\Delta)[\partial_i\vec{S}(\mathbf{r})]^2.
\end{equation}
We introduce a canting field $\vec{\chi}(\mathbf{r})$,
$\vec{\chi}^2(\mathbf{r})\le1$, perpendicular to a collinear ground
state $\vec{S}_0(\mathbf{r})\equiv\vec{S}_0$ of the pure system via
\begin{equation}\label{cafield}
  \vec{S}(\mathbf{r})=\sqrt{1-\vec{\chi}^2(\mathbf{r})}
  \vec{S}_0+\vec{\chi}(\mathbf{r}). 
\end{equation}
To show the canting instability we insert (\ref{cafield}) into the
Hamiltonian and minimize the energy with respect to the canting field
$\vec{\chi}$ after expanding the Hamiltonian up to quadratic order in
this field. In Fourier space the saddle-point equation reads
\begin{equation}\label{saddle}
  \vec{\chi}_{\mathbf{q}}=\Delta \frac{q_j}{\mathbf{q}^2}
  \int_{\mathbf{k}}k_j\vec{\chi}_{\mathbf{k}}.
\end{equation}
Multiplying equation (\ref{saddle}) with $q_i$ and integrating over $\mathbf{q}$ we get
\begin{equation}
  \vec{\mu}_i=\frac{\Delta}{d}\delta_{ij}\vec{\mu}_j,
\end{equation}
with
\begin{equation}
  \vec{\mu}_i:=\int_{\mathbf{k}}k_i\vec{\chi}_{\mathbf{k}}.
\end{equation}
This self-consistency condition on $\vec{\mu}_i$ implies
\begin{equation}
  \Delta=\Delta_{\mathrm{single}}=d
\end{equation}
for a non-vanishing solution of the saddle-point equation (\ref{saddle}).



\begin{thebibliography}{40}
\bibitem[{not({\natexlab{a}})}]{note:struct}
\bibinfo{note}{We refer to the square lattice only in a topological sense but
  not in the strict geometric sense. The materials of interest may exist in
  tetragonal and in orthorhombic phases. In the latter one the squares are
  deformed leading to a reduced symmetry. We neglect a possible anisotropy in
  the exchange couplings as well as couplings between more distant spins.}
\end{thebibliography}

\end{document}